\documentclass[aps,prb,twocolumn,superscriptaddress,showpacs]{revtex4-1}
\usepackage[unicode=true,pdfusetitle, bookmarks=false, breaklinks=false,pdfborder={0 0 0},backref=false,colorlinks=false] {hyperref}

\usepackage{graphicx}
\usepackage{epsfig}
\usepackage{subfigure}
\usepackage{setspace}
\usepackage{amsmath}

\begin{document}

\title{Polarization degenerate solid-state cavity QED}

\author{Morten P. Bakker}
\affiliation{Huygens-Kamerlingh Onnes Laboratory, Leiden University, P.O. Box 9504, 2300 RA Leiden, The Netherlands}
\author{Ajit V. Barve}
\affiliation{Departments of Electrical and Computer Engineering, University of California Santa Barbara, Santa Barbara, California 93106, USA}
\author{Thomas Ruytenberg}
\affiliation{Huygens-Kamerlingh Onnes Laboratory, Leiden University, P.O. Box 9504, 2300 RA Leiden, The Netherlands}
\author{Wolfgang L\"{o}ffler}
\affiliation{Huygens-Kamerlingh Onnes Laboratory, Leiden University, P.O. Box 9504, 2300 RA Leiden, The Netherlands}
\author{Larry A. Coldren}
\affiliation{Departments of Electrical and Computer Engineering, University of California Santa Barbara, Santa Barbara, California 93106, USA}
\author{Dirk Bouwmeester}
\affiliation{Huygens-Kamerlingh Onnes Laboratory, Leiden University, P.O. Box 9504, 2300 RA Leiden, The Netherlands}
\affiliation{Department of Physics, University of California Santa Barbara, Santa Barbara, California 93106, USA}
\author{Martin P. van Exter}
\affiliation{Huygens-Kamerlingh Onnes Laboratory, Leiden University, P.O. Box 9504, 2300 RA Leiden, The Netherlands}

\date{\today}

\begin{abstract}
A polarization degenerate microcavity containing charge-controlled quantum dots (QDs) enables equal coupling of all polarization degrees of freedom of light to the cavity QED system, which we explore through resonant laser spectroscopy.
We first measure interference of the two fine-split neutral QD transitions and find very good agreement of this V-type three-level system with a coherent polarization dependent cavity QED model. We also study a charged QD that suffers from decoherence, and find also in this case that availability of the full polarization degrees of freedom is crucial to reveal the dynamics of the QD transitions.
Our results pave the way for postselection-free quantum devices based on electron spin--photon polarization entanglement.

\end{abstract}

\maketitle
\section{Introduction}
Quantum dots (QDs) embedded inside microcavities are of interest for hybrid optical-solid-state quantum information schemes \cite{Waks2009,Cirac1999}, and long-distance quantum networks \cite{Yao2005,Kimble2008}.
A key ingredient is the realization of entanglement between a QD-spin and a single photon.
Several experiments have demonstrated this by utilizing spontaneous emission \cite{DeGreve2012,Gao2012,Schaibley2013}, but these methods require postselection and are therefore not suitable for deterministic approaches.
The need for postselection can be eliminated by using the spin-dependent reflection or transmission of a photon by a quantum dot in a cavity QED system.
Several protocols have been proposed that either require polarization degenerate microcavities in order to couple with circular polarized light \cite{Hu2009,Bonato2010}, or would be aided in order to match more easily with linear polarized transitions \cite{Sun2014}.
Further key system requirements are charge controlled QDs and access to the Purcell or strong coupling regimes, which has been realized in photonic crystal cavities \cite{Carter2013} and micropillars \cite{Rakher2009}.
Micropillars have the additional benefit of mode-matching to external fields and polarization control of the cavity modes \cite{Stoltz2005,Strauf2007,Dousse2010,Reitzenstein2010,Loo2012,Nowak2013}.

\begin{figure}[b]
\centering
\centerline{\includegraphics[angle=0]{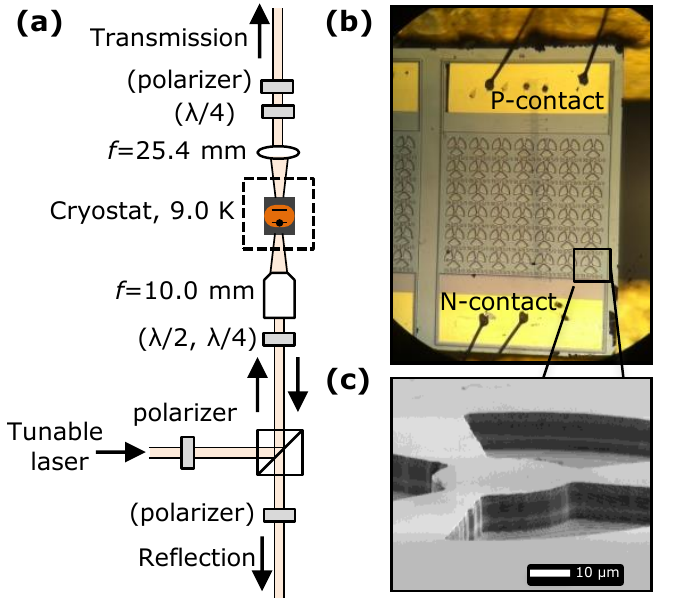}}
\caption{(a) Schematic of the setup. Light is coupled into a microcavity mode and the reflection and transmission spectra are recorded using single-photon avalanche photodiodes. The elements with names between brackets can be introduced for polarization analysis with either linear or circularly polarized light. $\lambda/2$ ($\lambda/4$): half- (quarter-) waveplate.
(b) Optical microscope image of a sample and (c) Electron micrograph of the cavity region.}
\label{Fig1}
\end{figure}

In this letter we report on a system exhibiting all these features, being a charge controlled quantum dot coupled to a polarization degenerate micropillar cavity.
The microcavity consist of two distributed Bragg reflectors, a $3/4\lambda$ thick aperture region for transverse mode confinement, and a $\lambda$ thick cavity layer, containing InAs self-assembled QDs embedded inside a PIN-diode structure \cite{petroff2001,Stoltz2005}.
By systematically varying the size and shape of the oxide aperture, we were able to select on average one polarization degenerate cavity (polarization splitting $<$3 GHz) out of an ($6\times7$) array \cite{bakker2014}.
This technique could be combined with a technique to actively tune the polarization properties by applying laser-induced surface defects \cite{Bonato2009}, to enhance the sample yield.
We tune the QD transition through the cavity resonance by the quantum confined Stark effect, induced by an applied bias voltage across the active region \cite{fry2000b,Warburton2000}.
In principle this can be combined with other QD tuning techniques, such as strain tuning \cite{Seidl2006,Sun2013,Gudat2011}, which would further increase the sample yield.
Further details on the sample structure and characterization can be found in the Appendix \ref{A}.
The setup, an optical, and an electron microscope image of the sample are shown in Fig.~\ref{Fig1}.

This system enables polarization resolved studies, which, as we will demonstrate, provides insight in the excitonic coherence of the system.
First we study the coherent interaction of charge-neutral quantum dot transitions with resonant laser light and give a theoretical description.
Then we investigate a singly charged QD and study its more complex dynamics, which we can describe with a second, decoherent model where all spin-photon entanglement is lost.

\begin{figure}[b!]
\centering
\centerline{\includegraphics[angle=0]{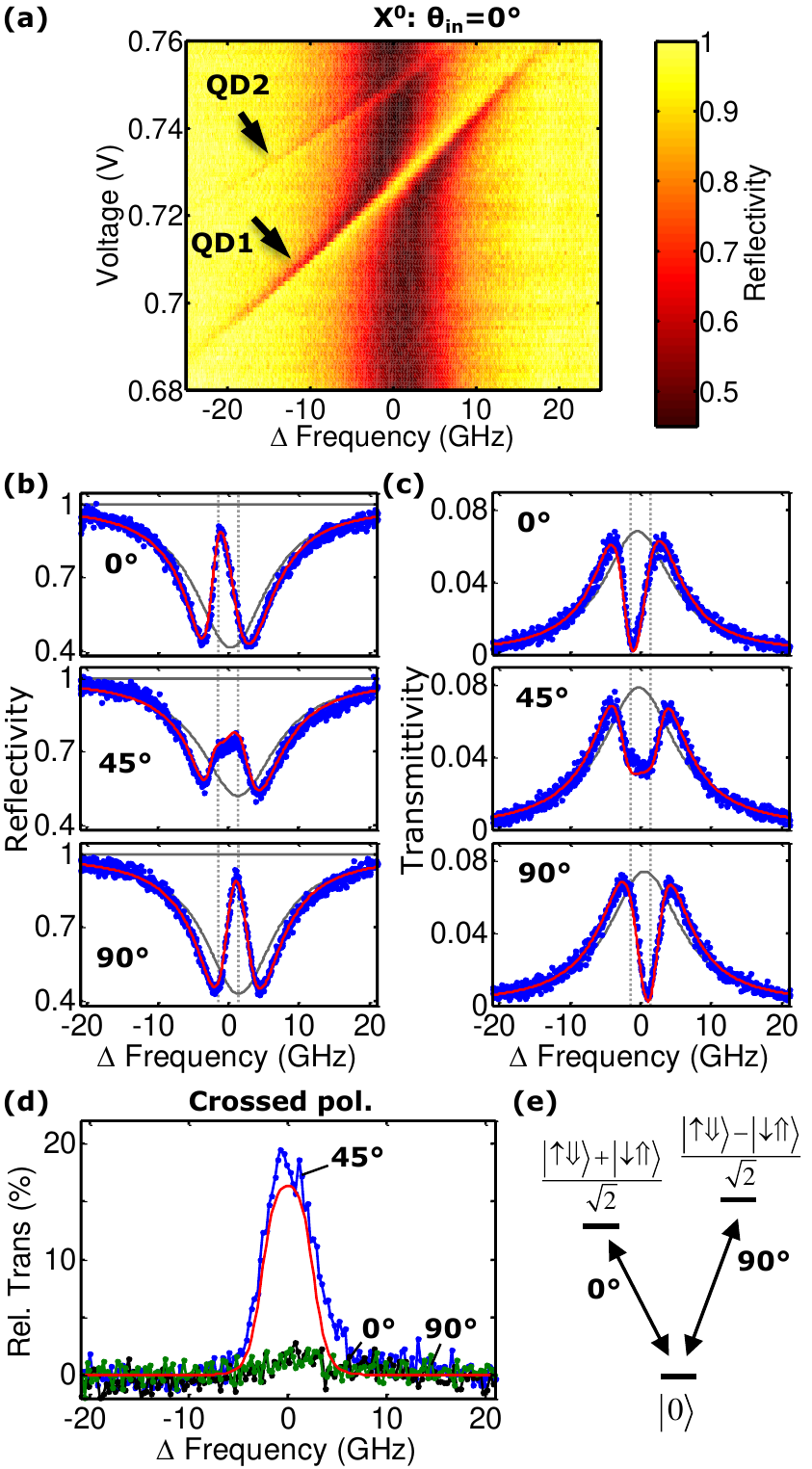}}
\caption{(a) Reflectivity measurement of two neutral QDs as function of the scanning laser frequency and applied voltage. The incoming polarization $\theta_{in} = 0^{\circ}$, $P_{laser} = 1$ pW and $\lambda \approx 940$ nm.
Panel (b,~c) show reflectivity and transmittivity spectra of QD1 recorded at $V = 0.725$ V for various incoming linear polarizations.
Blue points: experimental data. Red line: fitted curve using Eqs. \ref{Eq1} and \ref{Eq2}. Grey curve: empty cavity, calculated from the fits. Vertical dashed lines: frequencies corresponding to the two fine-split transitions. (d) Transmittivity spectra when a crossed polarizer is used with respect to the incoming polarization, relative to the maximum transmittivity of an uncoupled cavity. The red line is calculated using Eqs. \ref{Eq1} and \ref{Eq2} and the parameters obtained from the fits in (b,~c). (e) Energy level diagram of the ground-state and lowest energy excited states of a neutral QD.}
\label{Fig2}
\end{figure}

\section{Neutral quantum dot}
The lowest energy levels of a neutral QD are depicted in Fig.~\ref{Fig2}~(e).
Due to the QD anisotropy, the electron-hole exchange interaction leads to a fine-structure splitting of the excited states ($\sim3$ GHz for the QD under study), the neutral ground state is coupled to two excited states by two linear orthogonally polarized transitions.
In the resonant reflection measurements in Fig.~\ref{Fig2}~(a), the QD-cavity anti-crossing, as a hallmark of strong to intermediate QD-cavity coupling, is clearly visible.
Low laser power ($P_{laser} = 1$ pW) is used in order to avoid saturation of the QD transition, charging \cite{Houel2012} and dynamical nuclear spin polarization effects \cite{Urbaszek2013}.
Fig.~\ref{Fig2}~(b,~c) show reflection and transmission spectra for a voltage $V$=0.725 V, where QD1 is tuned into resonance with the cavity.
The spectra are recorded for three linear polarizations that couple either with the low frequency QD transition ($\theta_{in}=0 ^{\circ}$), or the high frequency QD transition ($\theta_{in}=90 ^{\circ}$), or both QD transitions ($\theta_{in}=45 ^{\circ}$).

For $0^{\circ}$ and $90^{\circ}$ polarization we observe that the quantum dot is able to restore high cavity reflectivity with near-unity fidelity, but this effect appears to be reduced for 45$^\circ$.
Additionally we show spectra when a crossed polarizer is used in the transmission path in Fig.~\ref{Fig2}~(d).
We see that for $0^{\circ}$ and $90^{\circ}$ the light matches the natural polarization axes of the QD and that this polarization is maintained, resulting in a very low signal.
For $45^{\circ}$ incoming polarization the transmission is significant however.
In the following, we develop a theoretical model to gain insight into the dynamics.

The transmission amplitude through a cavity with a coupled two-level system is given by \cite{Loo2012,auffeves2007,waks2006}:
\begin{equation}\label{Eq1}
t = \eta_{out}\frac{1}{1-i\Delta+\frac{2C}{1-i\Delta'}},
\end{equation}
where $\Delta = 2(\omega-\omega_c)/\kappa$ is the relative detuning between the laser ($\omega$) and cavity ($\omega_c$) angular frequencies, $\Delta' = (\omega-\omega_{QD})/\gamma_\bot$ is the relative detuning between the laser and QD transition ($\omega_{QD}$) and $\eta_{out}$ is the output coupling efficiency.
The device cooperativity is $C=g^2/\kappa\gamma_\bot$ where, $\kappa$ is the total intensity damping of the cavity, $\gamma_\bot$ is the QD dephasing rate and $g$ is the QD-mode coupling strength.
We obtain close to perfect mode-matching, and therefore the total transmittivity through the cavity is given by $T = |t|^2$, and the total reflectivity is given by $R = |1-t|^2$.
A more detailed description of Eq. (\ref{Eq1}) is provided in Appendix \ref{B}.

An important figure of merit of the QD-cavity system is the cooperativity parameter $C$.
By fitting our model to the experimental data in Fig.~\ref{Fig2} for $\theta_{in}=0^{\circ}$ and $\theta_{in}=90^{\circ}$, we find $C = 2.5\pm 0.5$, a value similar to previously reported \cite{Loo2012}.
We also obtain $\gamma_\bot = 2.0 \pm 0.5$ ns$^{-1}$, which corresponds to a total dephasing time $\tau = 500$ ps, and total cavity damping rate $\kappa=77$ ns$^{-1}$, which corresponds to a quality factor of $Q\sim 2.6*10^4$, see Appendix \ref{B}.
Since $\gamma_\bot < 2g=39$ ns$^{-1}<\kappa$, this places the system in the intermediate coupling regime.

The lineshapes corresponding to an empty cavity can be calculated from the fitted curves and are shown by the grey curves in Fig.~\ref{Fig2}~(b,c).
The very small dependence of the cavity resonance frequency on the polarization angle confirms the high degree of polarization isotropy of this device.

To account for the fine-structure splitting of the neutral QD transitions in the polarization-degenerate cavity, we write the transmission of the system in terms of a Jones matrix $\textbf{t}(\omega) = \begin{pmatrix} t_x(\omega) &0\\0 & t_y(\omega) \end{pmatrix}$.
The measured transmittivity therefore depends on the input and output polarization as
\begin{equation}\label{Eq2}
    t_{\theta_{out},\theta_{in}}(\omega) = \textbf{e}^{\dagger}_{out}\textbf{t}(\omega) \textbf{e}_{in},
\end{equation}
where $\boldsymbol{e_{i}} = (cos(\theta_{i}),sin(\theta_{i}))$ defines the linear input/output ($i = $in/out) polarization with angle $\theta_i$.
This model assumes that when the two transitions are excited simultaneously ($\theta_{in} = 45^\circ$), coherence in the system is fully maintained leading to quantum interference between the transmission amplitudes $t_x$ and $t_y$.
In an incoherent system we would obtain a classical mixture of the excited states, making such interference impossible.
The reflectivity is calculated in a similar way by using $r_{x/y} = 1-t_{x/y}(\omega)$ in the Jones matrix.

\begin{figure}[ht!]
\centering
\centerline{\includegraphics[angle=0]{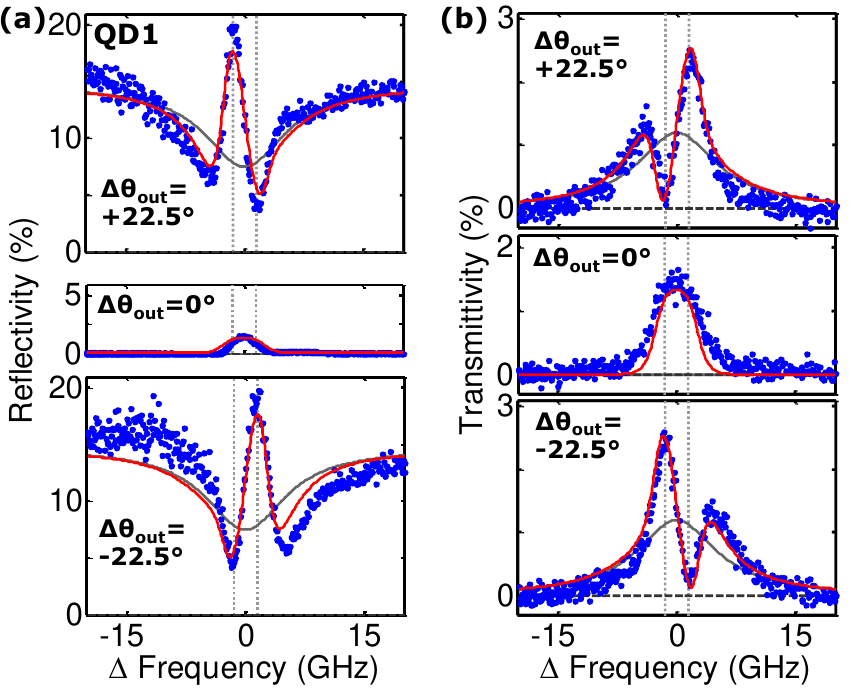}}
\caption{Resonant (a) reflection and (b) transmission spectroscopy with a neutral QD (QD1 in Fig.~\ref{Fig2}) for $\theta_{in} = 45 ^{\circ}$ and for various $\theta_{out} = \theta_{in} + 90^{\circ} + \Delta\theta_{out}$. Blue dots: experimental data. Red lines: predicted curves using Eqs. \ref{Eq1} and \ref{Eq2} and the parameters obtained from the fits in Fig.~\ref{Fig2}~(b,c). Grey lines: predicted curves corresponding to an empty cavity. Vertical dashed lines mark the two transitions split by the fine-structure interaction.
}
\label{Fig3}
\end{figure}

To further explore the validity of Eq.~(\ref{Eq2}) and to demonstrate the full power of polarization degenerate cavity QED, we show in Fig.~\ref{Fig3}~(a,~b) reflection and transmission spectra for $\theta_{in} = 45^{\circ}$, while $\theta_{out} = \theta_{in} + 90^{\circ} + \Delta\theta_{out}$ is varied.
For $\Delta\theta_{out} = 0^{\circ}$, the crossed polarizer condition, the transmission and reflection spectra consist of two partially overlapping Lorentzian lines split by $\sim$3 GHz.
The phase difference between these two resonances becomes apparent for the $\Delta\theta_{out} = +22.5^{\circ}$ ($-22.5^{\circ}$) spectra, which can be seen as the \emph{coherent sum} of the $\Delta\theta_{out} = 0^{\circ}$ and the $\Delta\theta_{out} = +45^{\circ}$ ($-45^{\circ}$) spectra, where the latter only contains the high (low) frequency transition.
All the red curves in Fig.~\ref{Fig2} and \ref{Fig3} are produced with the same parameters for $C$, $\kappa$ and $\gamma_\bot$ and fit the experimental data very well.
The results demonstrate how in a polarization degenerate cavity the fine-split excited states of a neutral QD can be simultaneously addressed in a coherent way.
Furthermore, these interference measurements hold great promise as a clever combination of $\textbf{e}_{in}$ and $\textbf{e}_{out}$ can be used to tune the constructive or destructive interference between $t_x$ and $t_y$.
This forms a generic technique to increase the ratio between an uncoupled and a coupled cavity system, and thereby the fidelity of entanglement operations.

\section{Singly charged quantum dot}
\begin{figure}[h!]
\centering
\centerline{\includegraphics[angle=0]{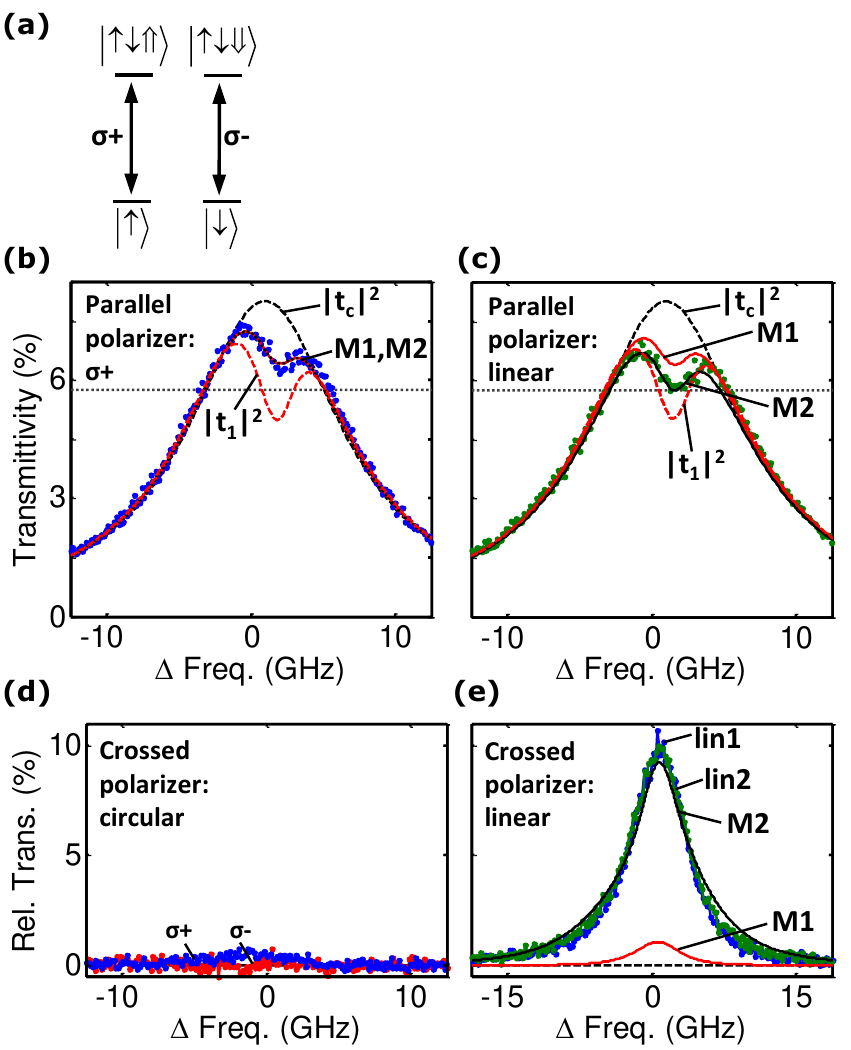}}
\caption{(a) Energy-level diagram of a singly charged QD.
Transmission spectra for $P_{laser} = 10$ pW are shown for circular and linear polarization, analyzed with a (b) and (c) parallel or (d) and (e) crossed polarizer. The red-black dashed line in (b) is a fit of Eq.~(\ref{Eq6}) (coherent model, M1) to the data, which yields the same result as Eq.~(\ref{Eq8}) (decoherent model, M2).
The red (black) solid lines in (c) and (e) predict the experimental data using Eq.~(\ref{Eq6}) [Eq.~(\ref{Eq8})].
Black (red) dashed curves: empty (coupled) cavity calculations.
}
\label{Fig4}
\end{figure}

Now we turn to a different QD in the same polarization degenerate cavity, but operated in a voltage regime around $0.9$ V where it is singly negatively charged.
This system is of particular importance in quantum information as the optical transitions are polarization degenerate (see Fig.~\ref{Fig4}~(a)), due to cancellation of electron-hole exchange interaction, and enables coherent control of the resident electron spin if a small in-plane magnetic field is applied.
We first focus on Fig.~\ref{Fig4}~(b,~c), which shows transmission spectra when circularly ($\sigma^+$) or linearly polarized light is coupled into the cavity and transmitted light of the same (i.e., parallel) polarization is recorded.
We define the contrast as $(|t_c|^2-T)/|t_c|^2$, with the measured transmittivity $T$ with a QD and the calculated transmittivity $|t_c|^2$ without a QD.
While for the neutral QD case we found contrasts of $>91\%$ in Fig.~\ref{Fig2}~(c), we now observe a strongly reduced contrast of the QD resonance, which is $\sim19\%$ when circularly polarized light is used and $\sim26\%$ for linear polarization.

We use a slightly larger laser power ($P_{laser} = 10$ pW) compared to the neutral QD as we find that the charging effects are now significantly smaller, due to less absorption of the resonant laser at this voltage.
Furthermore, this intensity corresponds to a mean intracavity photon number $\langle\bar{n}\rangle = |t|^2 P_{laser}/(\kappa_m\hbar\omega) < 0.001 $, and is therefore sufficiently small to prevent QD saturation effects from occurring.

In addition, we compared the cross-polarized transmitted intensity for circular and linear polarized light.
For circular ($\sigma^+$ and $\sigma^-$) polarization, shown in Fig.~\ref{Fig4}~(d), we observe negligible transmission, indicating that circular polarization remains unchanged.
Surprisingly, for two linear orthogonal (lin1 and lin2) polarizations displayed in Fig.~\ref{Fig4}~(e), we observe that about 10\% of the light is transmitted relative to $|t_c|^2$, despite the low cooperativity (see below).

We will first try to explain our observations with a coherent model, which we adapt to the four-level system of a charged QD shown in Fig.~\ref{Fig4}~(a):
The ground state consists of the two spin eigenstates, oriented in the out-of-plane direction, which couple with two corresponding trion lowest-energy excited states by degenerate circularly polarized optical transitions carrying spin $\sigma^{\pm} = \pm 1$.
We write $t_1^{\pm}\equiv t_1$ for the corresponding transmission amplitudes of $\sigma^{\pm}$ polarized light coupling with a corresponding transition, and $t_c^{\pm}\equiv t_c$ for the case of an empty cavity.
Since we do not control the electron spin state it can be in any random state $|\phi_{spin}\rangle = \alpha|\uparrow\rangle + \beta|\downarrow\rangle$.
With the incoming photon state $|\phi_{in}\rangle = \gamma|+\rangle + \delta|-\rangle$, we obtain for the input quantum state $|\Psi_{in}\rangle = |\phi_{in}\rangle\otimes|\phi_{spin}\rangle$.
The spin-selective interaction with the cavity-QD system entangles the photon with the electron spin via
\begin{equation}\label{Eq4}
    |\Psi_{out}\rangle = t_1\gamma\alpha|+\uparrow\rangle+t_c\gamma\beta|+\downarrow\rangle+t_c\delta\alpha|-\uparrow\rangle+t_1\delta\beta|-\downarrow\rangle.
\end{equation}

We then project this output state onto the detected polarization $|\phi_{out}\rangle = \gamma'|+\rangle + \delta'|-\rangle$, and take the trace over the electron spin to obtain the projected transmission:
\begin{equation}\label{Eq6}\begin{split}
T = |t_1\gamma\gamma' + t_c\delta\delta'|^2|\alpha|^2 + |t_c\gamma\gamma' + t_1\delta\delta'|^2|\beta|^2.
\end{split}
\end{equation}

Since we do not control the spin state we use $|\alpha|^2=|\beta|^2=0.5$ for the balanced case.
Note that this model (M1) is coherent in the sense that it still contains interference between the $t_1$ and $t_c$ terms.

The red solid line in Fig.~\ref{Fig4}~(c) shows how model M1 fits our data for the optimum cavity-QD coupling and QD dephasing parameters $C=0.13$ and $\gamma_\bot = 9.5$ ns$^{-1}$.
The dephasing rate can not be explained by the decay rate of the excited state, since lifetime measurements showed this to be about 1.2 ns.
Instead, we attribute this much faster dephasing rate to an efficient cotunneling process across the 20 nm electron tunnel barrier, which is expected to be more pronounced for the flatter conduction band here compared to the neutral QD case presented before.
This fast dephasing also reduces the cooperativity, which, however, might also be reduced due to low spatial overlap between the QD and the cavity mode.
We expect that utilizing a thicker 35 nm tunnel barrier will decrease the cotunneling process and enable high fidelity spin state preparation \cite{Atature2006}.

Next we consider the linear-polarization data shown in Fig.~\ref{Fig4}~(c,e), where the model prediction is shown by red lines.
Eq.~(\ref{Eq6}) predicts that purely circular polarized light should pass the cavity unmodified, and can therefore be fully blocked by a crossed polarizer ($\gamma\gamma'=\delta\delta'=0$), which is indeed what we observe experimentally in Fig.~\ref{Fig4}~(d).
Significant discrepancies between the data and our model are however observed in Fig.~\ref{Fig4}~(c) and in (e) particularly, where the cross-polarized transmission signal for linear-polarizations lin1 and lin2 is much larger than expected.
This can not be caused by an energy splitting, or phase difference, between the two transitions, as these splittings would have been visible in the data.
Furthermore, the observed cross-polarized transmission is so large that it would require $C>0.8$ in Eq.~(\ref{Eq6}) to explain the cross-polarized transmission in Fig.~\ref{Fig4}~(e), while we found $C=0.13$ for the fit in Fig.~\ref{Fig4}~(b).

This result therefore indicates that additional dephasing processes take place that project linear polarized light on the preferred circular basis of the QD transitions.
The preference for this basis is known from literature Refs. \cite{Hogele2005,Atature2006,Warburton2013} and is experimentally demonstrated by the fact that circular polarized light remains circular polarized after the interaction with the QD--cavity system.
If the absorption and re-emission of linear light would be a fully coherent process, the linear polarization should largely remain, which is clearly not the case in Fig.~\ref{Fig4}~(e).

To model the results, we now introduce a tentative model (M2) that describes the spin-exciton system as if it were fully decoherent, meaning that any light interacting with the QD is instantaneously projected on the QD transition polarization basis.
This corresponds to immediate decoherence of the entangled state described by Eq.~(\ref{Eq4}) and elimination of interference between the $t_1$ and $t_c$ terms in Eq.~(\ref{Eq6}).
Since only a fraction of the light that enters the cavity becomes entangled with the QD spin state, we first need to calculate the fraction of the light that did not interact.
We estimate this fraction $T_0$ by multiplying the cavity transmission with the QD response function: $T_0 = |t_c|^2\times|\frac{1}{1+\frac{2C}{1-i\Delta'}}|^2$.
The intensities of the circularly polarized components of the transmitted light that interacted with a parallel or opposite electron spin are now given by $T_1' = |t_1|^2-T_0$ and $T_c' = |t_c|^2-T_0$, respectively.

The total transmitted intensity corresponds now to the incoherent sum of five transmission channels:
\begin{equation}\label{Eq8}\begin{split}
    T = T_0 |\langle\phi_{out}|\phi_{in}\rangle|^2+T_1'|\gamma\alpha\langle\phi_{out}|+\rangle|^2+T_c'|\delta\alpha\langle\phi_{out}|-\rangle|^2\\
    +T_c'|\gamma\beta\langle\phi_{out}|+\rangle|^2+T_1'|\delta\beta\langle\phi_{out}|-\rangle|^2.
\end{split}
\end{equation}

The transmission predicted by the incoherent model (M2, Eq.~(\ref{Eq8})) and coherent model (M1, Eq.~(\ref{Eq6})) are equivalent in case of circular incoming polarization (Fig.~\ref{Fig4}~(b,d)).
They differentiate however in case of the linear-polarization data in Fig.~\ref{Fig4}~(c,e).
The solid black curves predicted by the incoherent model (M2), based on the parameters deduced from Fig.~\ref{Fig4}~(b), agrees very well while the coherent model (M1) does not.

While the polarization degenerate microcavities enables systematic polarization analysis and the identification of a high degree of decoherence in the charged QD system, the exact origin of decoherence is not known to us.
We think it is related to the cotunneling process and future sample designs with thicker tunnel barriers will resolve this issue.

\section{Conclusion}
In conclusion, we have demonstrated a polarization degenerate solid-state cavity QED system with charge control, which allows full use of all polarization degrees of freedom.
Here, simple polarimetric reflection and transmission measurements enable the study of the coherence properties of the coupled QD--cavity system, for neutral and charged quantum dots.
This is an important advance for fundamental studies of spin dynamics and optical interactions in solid-state cavity QED systems, and an important step towards quantum information applications with single electron and hole spin qubits, and postselection-free spin--photon polarization interaction.

\begin{acknowledgements}
We would like to thank Alan Zhan for help with the sample characterization measurements.
This work was supported by NSF under Grant No. 0960331 and 0901886 and FOM-NWO Grant No. 08QIP6-2.
\end{acknowledgements}

\appendix
\section{Sample structure and characterization}
\label{A}
The sample under study has been grown by molecular beam epitaxy on a GaAs [100] substrate.
Two distributed Bragg reflectors (DBR) surround an aperture region and a $\lambda$ thick cavity region containing in the center InAs self-assembled quantum dots (QDs).
The top DBR mirror consists of 26 pairs of $\lambda/4$ layers of GaAs and Al$_{0.90}$Ga$_{0.10}$As, while
the bottom mirror consists of 13 pairs of layers of GaAs and AlAs and 16 pairs of GaAs and Al$_{0.90}$Ga$_{0.10}$As layers.
This way the reflectivities of top and bottom mirrors are matched in order to enable transmission and reflection measurements and optimize the incoupling efficiency.
The oxidation aperture consists of a 10 nm AlAs layer embedded between 95 nm Al$_{0.83}$Ga$_{0.17}$As and 66 nm Al$_{0.75}$Ga$_{0.25}$As layers, providing a linearly tapered oxidation upon wet oxidation.
The QDs are separated by a 20 nm GaAs tunnel barrier to n-doped GaAs (Si dopant, concentration $2.0\times10^{18}$ cm$^{-3}$) and by a 107 nm GaAs layer to p-doped GaAs (C doping, concentration $1.0\times10^{18}$ cm$^{-3}$).

By analyzing the confined optical modes and the wavelength splitting between the fundamental and first order optical modes, an estimation can be made of the maximum Purcell factor and the numerical aperture (NA) of the fundamental mode.
A high Purcell factor is necessary to observe QD couplings close to the strong coupling regime, while a modest NA enables perfect mode-matching to external fields.

\begin{figure}[h]
\centering
\centerline{\includegraphics[angle=0]{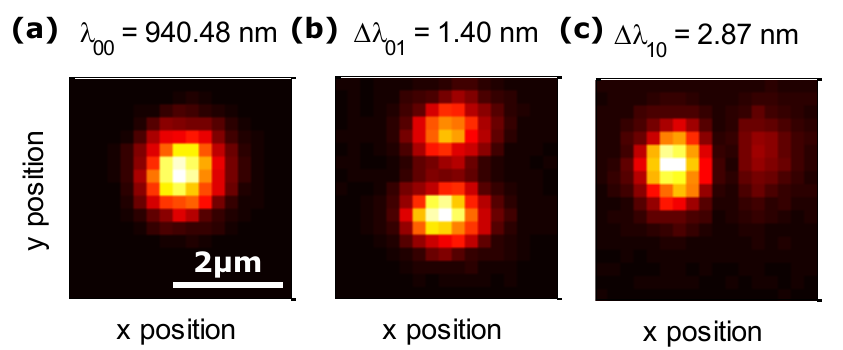}}
\caption{Spatial PL scans of the Hermite-Gaussian modes, where (a) is the fundamental mode $\Psi_{00}$ and (b,c) the first order $\Psi_{10}/\Psi_{01}$ modes. Light: more PL counts. The captions denote the wavelength $\lambda_{00}$ of the fundamental mode, or the wavelength splitting $\Delta\lambda_{10/01} = \lambda_{00}-\lambda_{10/01}$.}
\label{Suppl_Fig1}
\end{figure}

To characterize the optical properties of the confined modes, the sample is excited using an 852 nm laser diode and photoluminescence as function of position is recorded using a spectrometer.
Hermite-Gaussian modes are clearly identified in Fig. \ref{Suppl_Fig1}.
Following methods described in \cite{Bonato2012} we calculate the mode volume $V$ using:
\begin{equation}\label{Suppeq1}
    V = L_{cav}\frac{\lambda_{00}^3}{8\pi n_0^2\sqrt{\Delta\lambda_{01}\Delta\lambda_{10}}},
\end{equation}
where $L_{eff} \approx 5\lambda_{00}/n\approx 1.4$ $\mu$m is the effective cavity length, $\lambda_{00} = 940.48$ nm is the wavelength of the fundamental mode in vacuum, $n \approx 3.25$ is the average refractive index, and $\Delta\lambda_{01/10}$ are the mode splittings between the $\Psi_{01/10}$ modes and the $\Psi_{00}$ mode.
Filling in the experimentally obtained values for the modesplitting, we obtain $V = 2.2$ $\mu$m$^3$.
The expected maximum Purcell factor $P$ is given by:
\begin{equation}\label{Suppeq2}
    P=\frac{3}{4\pi^2}(\frac{\lambda_{00}}{n_0})^3\frac{Q}{V},
\end{equation}
where $Q = 2.6*10^4$ is the quality factor measured during the resonant spectroscopy scans.
Using the above mentioned values we find $P = 22$. 
The intensity of the fundamental mode, perpendicular to the propagation direction $\hat{z}$, has the form: $I\propto \textrm{exp}[-2(\frac{x^2}{w_x^2} + \frac{y^2}{w_y^2})]$, where $w_{x/y} = \frac{1}{n_0 \pi}\sqrt{\frac{\lambda_{00}^3}{2 \Delta\lambda_{10/01}}}$ is the mode waist.
The numerical aperture of the Gaussian beam originating from the fundamental mode is given by NA$_{x/y} = \textrm{sin}(\frac{\lambda_{00}}{\pi W_{x/y}})$, which gives NA$_x = 0.18$ and NA$_y = 0.25$.
The NA of the used objective 0.4, enabling perfect mode-matching.

\section{Complete description of the transmission amplitude}
\label{B}
The transmission amplitude through a cavity with a coupled QD is given by \cite{Loo2012,auffeves2007,waks2006}:
\begin{equation}\label{Eq1b}
t = \eta_{out}\frac{1}{1-i\Delta+\frac{2C}{1-i\Delta'}},
\end{equation}
where the parameters are defined in the main text.
We will here quantify the role of losses and its effect on the out-coupling efficiency $\eta_{out}=\frac{2\kappa_m}{\kappa}$, defined as the probability that a photon in the mode will leave the cavity through the top or bottom mirror.
Here $\kappa_m$ is the damping rate of each Bragg mirror, $\kappa_s$ is the scattering and absorption rate inside the cavity, and $\kappa = 2\kappa_m+\kappa_s$ is the total cavity intensity damping rate.
Furthermore $\kappa_m = T_{mirror}/t_{round}$, where $T_{mirror}$ is the transmittivity of a single mirror and $t_{round} = 2nL_{cav}/c$ is the cavity round trip time.
$n$ is the average refractive index, $L_{cav} \approx 5 \lambda/n$ is the effective cavity length, $c$ is the speed of light and $\lambda \approx 940$ nm is the wavelength in vacuum.

The mirror damping rate $\kappa_m \approx 11$ ns$^{-1}$ is calculated from the sample design parameters.
Three observations consistently yield $\kappa_s \approx 55$ ns$^{-1}$: (i) the measured quality factor $Q \approx 2.6\times10^4$ is lower than $Q=9.1\times10^4$ as determined by the mirror transmittivity $T_{mirror}= 3.4*10^{-4}$ and cavity length, and corresponds to $\kappa= 77$ ns$^{-1}$, (ii) the minimum reflectivity of the empty cavity $\frac{R_{min}}{R_{max}} = |1-\eta_{out}|^2 \approx 0.5$, and (iii) the maximum transmission $T_{max} = |\eta_{out}|^2 \approx0.08$, (not taking into account a $\sim 30\%$ reflectivity at the GaAs to air interface at the back of the sample).
We attribute this scattering rate $\kappa_s$ to (spectrally broad) absorption losses in the doped layers and scattering by the oxide aperture.
Reducing $\kappa_s$, for example by using a lower doping concentration, is a major concern in future sample designs.

Finally we will comment on the case of non-perfect mode matching.
The total transmission $T$ through the cavity is then given by $T = \eta_{in}\eta_{T} |t|^2$, where $\eta_{in}$ is the in-coupling efficiency and $\eta_{T}$ is the collection efficiency at the transmission port.
The total reflection is given by $R =\eta_{R}|1-\eta_{in}t|^2$, where $\eta_{R}$ is the collection efficiency at the reflection port.
In case of perfect mode matching $\eta_{in}=\eta_{R}=\eta_{T}=1$.

\end{document}